\documentclass[acmlarge,nonacm]{acmart}
\usepackage[utf8]{inputenc}
\usepackage[T1]{fontenc}
\usepackage[multiple]{footmisc}
\usepackage[font=itshape,indentfirst=false]{quoting}

\usepackage{breakurl}
\usepackage{float}
\usepackage{booktabs}
\usepackage{multirow}
\usepackage{icomma}
\usepackage{subfig}
\usepackage{hyphenat}
\usepackage{tabularx}

\usepackage{phdcode}
\usepackage{phdquery}
\usepackage{phdmisc}

\usepackage{graphicx}
\graphicspath{{figures/}}

\begin{document}

\title{Fatigued PageRank}

\author{José Devezas}
\email{jld@fe.up.pt}
\orcid{0000-0003-2780-2719}

\author{Sérgio Nunes}
\email{ssn@fe.up.pt}
\orcid{0000-0002-2693-988X}

\affiliation{%
  \institution{INESC TEC \& Faculty of Engineering, University of Porto}
  \streetaddress{Rua Dr. Roberto Frias, s/n}
  \postcode{4200-465}
  \city{Porto}
  \country{Portugal}}

\begin{abstract}
  Connections among entities are everywhere. From social media interactions to web page hyperlinks, networks are frequently used to represent such complex systems. Node ranking is a fundamental task that provides the strategy to identify central entities according to multiple criteria. Popular node ranking metrics include degree, closeness or betweenness centralities, as well as HITS authority or PageRank. In this work, we propose a novel node ranking metric, where we combine PageRank and the idea of node fatigue, in order to model a random explorer who wants to optimize coverage --- it gets fatigued and avoids previously visited nodes. We formalize and exemplify the computation of Fatigued PageRank, evaluating it as a node ranking metric, as well as query-independent evidence in ad hoc document retrieval. Based on the Simple English Wikipedia link graph with clickstream transitions from the English Wikipedia, we find that Fatigued PageRank is able to surpass both indegree and HITS authority, but only for the top ranking nodes. On the other hand, based on the TREC Washington Post Corpus, we were unable to outperform the BM25 baseline, obtaining similar performance for all graph-based metrics, except for indegree, which lowered GMAP and MAP, but increased NDCG@10 and P@10.
\end{abstract}


\maketitle

%***********************************************************************
\section{Introduction}
%***********************************************************************

The graph is a well-established data structure used to model a wide range of real-world relations, from social ties to protein-protein interactions, from co-authorship to the web graph, from flight patterns to time series, from term dependencies to entity relations. Network science is the area that lies between multiple domains, providing a common set of tools to study real-world networks. One of the fundamental tasks in the study of a network is the measurement of node importance or centrality, usually with the goal of obtaining an ordering or ranking. Node ranking has been a fundamental task, not only in network science, but also in information retrieval, where historical metrics like PageRank were used as query-independent evidence of the authority of a web page, in order to improve retrieval effectiveness. Many variants of PageRank have since then been developed, exploring aspects like different smoothing approaches, or applications that rely on contextual or visual features in addition to, or instead of, the traditional hyperlinks.

In this work, we propose a PageRank variant inspired by the analogy to neuronal fatigue, as briefly mentioned by von Neumann in his last lecture~\cite{Neumann2012}. In particular, he said that, after being stimulated and activated, a neuron will temporarily enter a state of fatigue, during which it will not respond. Taking this characteristic as a relevant element of cognition, we decided to explore a similar idea with random walks in a graph, proposing a PageRank application where nodes have a probability of getting fatigued, in which case they are excluded from a particular step of the walk. We propose that the indegree is used as an indicator of fatigue --- high indegree nodes have a higher probability of being visited and thus have a higher probability of getting fatigued. Accordingly, we combined the indegree with the transition matrix from PageRank, in order to obtain a Fatigued PageRank.

The remainder of this paper is organized as follows. In Section~\ref{sec:ref-work}, we cover relevant work, namely the idea of fatigue in neuroscience and computer science, as well as node ranking approaches. In Section~\ref{sec:random-explorer-model}, we suggest a random explorer model to illustrate the impact of fatigue in random walks, formalizing Fatigued PageRank and illustrating its computation through power iteration as an extension of PageRank. In Section~\ref{sec:eval:netsci}, we assess the effectiveness of Fatigued PageRank purely as a node centrality metric, when compared with baselines like the indegree, HITS authority or PageRank. Our evaluation is based on the Simple English Wikipedia link graph, extended with the transitions from the English Wikipedia clickstream, as a ground truth indicator of node importance. In Section~\ref{sec:eval:ir}, we explore the impact of Fatigued PageRank in information retrieval, when used as a query-independent feature, in combination with BM25, versus other graph-based metrics. The evaluation was carried over the link graph for TREC Washington Post Corpus, using the relevance judgments for TREC 2018 Common Core track. In Section~\ref{sec:concl} we present the final conclusions and propose two lines for future work.

%***********************************************************************
\section{Reference work}
\label{sec:ref-work}
%***********************************************************************

In this section, we introduce the idea of neuronal fatigue, as mentioned by Von Neumann, and describe its applications in computer science (Section~\ref{sec:ref-work:fatigue}). We then present an overview on node centrality metrics, covering random walk based approaches such as PageRank (Section~\ref{sec:ref-work:centralities}). Finally, we show how graph-based metrics can be used as query independence evidence of document importance during search, illustrating with two approaches for integration into the retrieval model (Section~\ref{sec:ref-work:search}).

%***********************************************************************
\subsection{Neuronal fatigue in computer science}
\label{sec:ref-work:fatigue}
%***********************************************************************

In preparation for Yale's Silliman Memorial Lectures\footnote{\url{https://en.wikipedia.org/wiki/Silliman_Memorial_Lectures}}, von Neumann highlighted the importance of jointly studying the computer and the brain~\cite{Neumann2012}, a work to be published posthumously for the first time in 1958. In fact, with his lecture, he provided enough common ground for crossover work between computer science and neuroscience, bringing the two areas closer together. One of the ideas studied in von Neumann's lecture was the fact that a neuron will become fatigued, for a period of time, after being stimulated:

\medskip

\begin{quoting}
  ``However, this is not the most significant way to define the reaction time of a neuron, when viewed as an active organ in a logical machine. The reason for this is that immediately after the stimulated pulse has become evident, the stimulated neuron has not yet reverted to its original, prestimulation condition. It is \textit{fatigued}, i.e.\ it could not immediately accept stimulation by another pulse and respond in the standard way. [\ldots] It should be noted that this recovery from fatigue is a gradual one [\ldots]''
\end{quoting}

\medskip

\noindent Despite the impact neuroscience has had in computer science, namely with neural networks, not many analogies using neuronal fatigue have been proposed. In fact, to our knowledge, only Xu and Yu~\cite{Xu2010} have used fatigue in the context of neural networks, as a part of a revised version of backpropagation for spam filtering. In this work, we introduce node fatigue as a signal used in the computation of PageRank. In Fatigued PageRank, node importance is measured by exploring the topology of the graph through random walks, while taking into account the probability of a node getting fatigued --- we call this the random explorer model.

%***********************************************************************
\subsection{Node ranking approaches}
\label{sec:ref-work:centralities}
%***********************************************************************

Node centrality measures the importance of a node in a graph according to a given criterion, be it the number of incoming links (indegree), the distance to other nodes (closeness~\cite{Bavelas1950}), the number of shortest paths that pass through the node (betweenness~\cite{Freeman1977}), or the influence of a node (eigenvector~\cite{Bonacich1972a,Bonacich1972b}). Eigenvector centrality, which is recursively computed based on the influence scores of neighboring nodes, has been particularly relevant in information retrieval, where it is frequently used as a relevance indicator to improve search. In particular, HITS~\cite{Kleinberg1999} (query-dependent) and PageRank~\cite{Page1999} (query-independent) are two of the more common variants of the eigenvector centrality used in information retrieval to rank nodes according to their influence.

HITS is an algorithm that provides an authority score, based on incoming links, and a hub score, based on outgoing links. HITS can be computed over any graph, however it is frequently computed over a query-dependent graph, built from a root set of pages that are relevant to the query. The root set is usually obtained based on a retrieval model like TF-IDF or BM25 and it is then expanded into a base set that includes all outgoing links and a subset of incoming links. While the number of outgoing links is usually small, the number of incoming links can be too high for an efficient computation. Thus, a parameter $d$ is used to define a ceiling for the number of incoming links to consider. When the number of incoming links surpasses $d$, then only a random sample of size $d$ is considered, otherwise all incoming links are considered. In its query-dependent application, HITS is more expensive to compute than PageRank, since it cannot be done offline. Like PageRank, HITS is also related to the leading eigenvector of a matrix derived from the adjacency matrix. Interestingly, the authority and hub scores are related to the leading eigenvectors of $AA^T$ and $A^TA$, respectively, both sharing the same eigenvalue~\cite[§3.2]{Saerens2005}.

PageRank~\cite{Brin1998,Page1999} is an elegant metric that offers multiple interpretations and computation approaches. It can be seen as the solution to a linear system~\cite{Gleich2004,Corso2005}, or as the eigenvector of the Markov chain derived from the graph --- after adding a teleportation term to the transition probabilities, in order to deal with sinks (i.e., pages without any links to other pages). It can be solved through Gaussian elimination, power iteration or even Monte Carlo methods~\cite{Avrachenkov2007}. Conceptually, PageRank is a random surfer model, where the probability of visiting a node reflects the behavior of a user that is randomly navigating the web by clicking hyperlinks, while occasionally jumping to a new page. This model is recursive, in the sense that it results in a centrality metric where the importance of a node depends on the importance of its neighbors --- the better connected a node is, either through quantity (i.e., many nodes) or quality (i.e., nodes that are themselves well connected), the higher the PageRank.

Researching PageRank has led to many applications of the metric~\cite{Gleich2015a}, exploring contextual information (e.g., Topic-Sensitive PageRank~\cite{Haveliwala2003}), combinations of features (e.g., Weighted PageRank~\cite{Dimitrov2017}), alternative smoothing approaches (e.g., Dirichlet PageRank~\cite{Wang2008}) or historical evidence (e.g., Multilinear PageRank~\cite{Gleich2015}). One of the variants, Reverse PageRank~\cite{Fogaras2003}, consists of simply reversing the edge direction and computing PageRank for this complementary graph. It is to PageRank what the hub score is to the authority score in HITS. Bar-Yossef and Mashiach~\cite{Bar-Yossef2008} have shown that Reverse PageRank is not only useful to select good seeds for TrustRank~\cite{Gyoengyi2004} and web crawling, but also, more interestingly, for capturing the semantic relatedness between concepts in a taxonomy. According to Gleich~\cite[§3.2]{Gleich2015a}, Reverse PageRank can be used to determine \emph{why} a node is important, as opposed to simply identifying \emph{which} nodes are important, as PageRank already does. While both metrics are individually useful, in this paper we present the initial steps as to understand whether a combination of both approaches can also be useful --- for HITS, positive results were achieved in the past for some topics, when combining the authority and hub scores~\cite[Table 3]{Aouadi2012}.

%***********************************************************************
\subsubsection{Random walks as an underlying tool for node ranking}
%***********************************************************************

A random walk is a succession of random steps that can be issued over a mathematical space, such as the Euclidean space (e.g., two dimensional steps of $-1$ or $+1$ over an x-axis and a y-axis). Random walks on graphs~\cite{Lovasz1993} follow a similar approach, but are constricted to the graph space, where a step consists of randomly selecting and following an outgoing edge to reach the corresponding node. The random surfer model proposed by PageRank can be seen as a random walk over a graph where teleportation is possible --- the corresponding Markov matrix is built from a mixture between the probabilities of navigation, given by the scaled adjacency matrix, and the probabilities of teleportation, usually uniform and over all nodes. There are, however, other PageRank variations that get inspiration from different types of random walks.

\sloppy Klopotek et al.~\cite{Klopotek2014} have explored PageRank computations based on lazy random walks, as well as random walks with back step (also covered by Sydow~\cite{Sydow2004}). While in a standard random walk a step is always taken, in a lazy random walk, there is a $50\%$ chance that the walker does not advance (it might stop to read the page). In random walks with back step, there is, instead, the chance that the walker goes back to a previous node (it might dislike the page and decides to go back to try another link). Klopotek et al.\ propose a boring factor to generalize the implementation of the three PageRank versions. While Sydow has shown that a random surfer with back step will result in a distinct PageRank, to our knowledge both the lazy and the back step PageRank versions lack evaluation as query-independent relevance scores in the context of search.

Multilinear PageRank~\cite{Gleich2015} was inspired by vertex-reinforced random walks \cite{Volkov2001}, where previously visited vertices are more likely to be visited in the future (their probability is reinforced). Random steps in Multilinear PageRank are more informed, as they are based on multiple previous states, instead of just one previous state. The idea of fatigue we introduce here is somewhat complementary, since, in our model, visited vertices are less likely to be visited in the future. However, for Fatigued PageRank, we approximate this, using the indegree as an indicator of fatigue, that is combined with the visitation probability during the preparation of the initial Markov matrix.

%***********************************************************************
\subsection{Using graph-based features to improve search}
\label{sec:ref-work:search}
%***********************************************************************

One way to combine graph-based features with classical information retrieval models is to use learning to rank, as illustrated for instance by Moreira et al.~\cite{Moreira2011}. In their work, they tackled the task of expert search over a DBLP test collection\footnote{DBLP-Citation-network V3: \url{https://aminer.org/citation}}\footnote{People Lists (Expert Lists): \url{https://aminer.org/lab-datasets/expertfinding/}}, combining text similarity features like BM25 with graph-based features like PageRank. They also explored graph-based features specific to bibliometrics, like the $h$-index and other more modern variants, as well as features based on author profiles.

We adopt a more classical approach for combining query-dependent and query-independent features, as described by Craswell et al.~\cite{Craswell2005} and conveniently available in Apache Lucene\footnote{\url{https://lucene.apache.org/}} since version 7.4.0. Craswell et al.\ explored reranking approaches, experimenting with BM25 as the baseline and PageRank as the query-independent evidence, while studying statistical dependence. PageRank was converted into a relevance weight and added to BM25 after applying one of three proposed functions: \emph{log}, \emph{satu} or \emph{sigm}. Blanco and Lioma~\cite[§5.2.5]{Blanco2012} have also used this type of integration for query-independent graph-based features, specifically exploring the \emph{satu} function. In this work, we use the \emph{sigm} function, with the parameters that generated the best MAP in the Craswell et al.\ experiments ($w = 1.8$, $k = 1$, $a = 0.6$).

%***********************************************************************
\section{Random explorer model}
\label{sec:random-explorer-model}
%***********************************************************************

Random walks have been at the core of centrality metrics like PageRank, which models the behavior of a random surfer in the web graph. This means that, for each random step, there is the probability that we follow a random outgoing edge. However, given the complementary event, we simply jump to a random node in the graph. In analogy to PageRank, we propose a random explorer model to motivate and describe Fatigued PageRank.

The goal of the random explorer is to survey the graph space by randomly traversing edges, as long as it avoids recently visited nodes in order to optimize coverage --- while the random surfer goes where the graph leads it, the random explorer actively tries to get to know the graph. We might say that the explorer gets fatigued and doesn't want to revisit nodes that have recently been considered --- why would an explorer want to go to places it has already seen, when there is still so much to discover? Similar to PageRank, the explorer can also get stuck in sinks or cycles, or even become so fatigued that there is no interesting unexplored edge to traverse, in which case the explorer teleports to a new location to continue the survey.

In the following sections, we will revise the computation of PageRank based on power iteration, introducing the notation we use and highlighting memory management via sparse matrix representations. We will then extend PageRank with fatigue and propose a computation approach for Fatigued PageRank.

%***********************************************************************
\subsection{From PageRank to Fatigued PageRank}
%***********************************************************************

The original PageRank corresponds to the stationary distribution of a Markov chain that models the transition probabilities between nodes in a web graph. Transitions can either happen through navigation (i.e., following a hyperlink) or through teleportation (i.e., randomly jumping to a web page). Fatigued PageRank considers similar navigation and teleportation behaviors, but introduces the concept of visitation fatigue. While in PageRank the user could only get bored and jump to another page, in Fatigued PageRank the user will also avoid recently visited pages (i.e., in the process of information seeking, the user will eventually get tired of revisiting a page and avoid it for a while). This means that memory is required to store fatigue information per node, eventually violating the Markov property (future states now depend on the current state and the fatigue state). In order to ensure that the Markov property is not violated, we approximate the probability of a node being fatigued based on its indegree --- the higher the number of incoming connections, the higher the probability that a node gets fatigued, and thus the lower the probability the node is visited due to fatigue.

%***********************************************************************
\subsubsection{A revision on PageRank computation using power iteration}
%***********************************************************************

In order to describe Fatigued PageRank, we will first illustrate the computation of PageRank based on power iteration, highlighting some simple techniques to minimize memory usage. Let us first assume a directed graph $G = (V, E)$ represented by its adjacency matrix $A$, where each row $i$ illustrates outgoing links from node $i$ to a node $j$. Based on $A$, we need to obtain a left stochastic matrix $S$ representing the outgoing transition probabilities (each column $j$ represents the probability of transitioning from node $j$ to a node $i$). As shown in Equation~\ref{eq:pr:H}, this can be partly done by normalizing each column, based on the sum of its elements. However, we must use a different strategy for dealing with columns that are all zeros (representing sink nodes, without outgoing edges), since we cannot divide by zero. In Equation~\ref{eq:pr:H}, we simply maintain the zero-sum columns, which means that $H$ still isn't stochastic (not all columns sum to one).

\begin{equation}
  \label{eq:pr:H}
  H_{ij} = \begin{cases}
    0 & \text{if } \sum_k (A^T)_{kj} = 0\\
    \frac{(A^T)_{ij}}{\sum_k (A^T)_{kj}} & \text{otherwise}\\
  \end{cases}
\end{equation}\\[-.5\baselineskip]

\noindent In order to obtain a stochastic matrix, despite possible sinks, we calculate matrix $S$ from matrix $H$ as described in Equation~\ref{eq:pr:S}.

\begin{equation}
  \label{eq:pr:S}
  S = H + \frac{1}{|V|} a e^T
\end{equation}\\[-.5\baselineskip]

\noindent This is done by first obtaining a binary vector $a$ that acts as a mask, where ones identify zero-sum columns. Using this mask, we then replace zero-sum columns by a uniform vector with the probability $\frac{1}{|V|}$ of randomly jumping to any node. The term $\frac{1}{|V|}a e^T$ is essentially a square matrix that repeats $\frac{1}{|V|}$ over all lines of zero-sum columns --- it's as if sinks are now linked to all nodes in the graph. Finally, using Jelinek-Mercer smoothing (linear interpolation), we combine $S$ with a teleportation term, obtaining the Markov matrix $\mathcal{M}$ that will be used in the power iteration to compute the PageRank vector $r$. Equation~\ref{eq:pr} illustrates the computation of PageRank based on a damping factor $\alpha$, which is usually set to $0.85$, the number of vertices $|V|$ and the column-vector $e$ of size $|V|$ and all ones. Power iteration is then initiated with any stochastic vector $r_t$, which is iteratively multiplied by $\mathcal{M}$, resulting in a normalized vector, until convergence --- i.e., until $r_{t+1} \approx r_t$, as determined by the L2-norm of the difference between $r_{t+1}$ and $r_t$ and an $\epsilon$ convergence constant, frequently set to $0.001$ or less.

\begin{equation}
  \label{eq:pr}
  \begin{split}
    r_{t+1} & = \frac{\mathcal{M} r_t}{\lVert\mathcal{M} r_t\rVert_1} = \mathcal{M} r_t\\
    \mathcal{M} & = \frac{1 - \alpha}{|V|} e e^T + \alpha S
  \end{split}
\end{equation}\\[-.5\baselineskip]

While PageRank can be computed using power iteration, as described in Equation~\ref{eq:pr}, it is easy, even for only a slightly large graph, to run out of memory during PageRank computation. This is because $\mathcal{M}$ is dense and, for dense matrices, space complexity is $O(|V|^2)$. So, for instance for a graph with 1 million nodes and assuming entries of 8 bytes, we would need $7.28$~TB of memory only to store $\mathcal{M}$, not even accounting for the overhead of the data structure. One way to mitigate this problem is to ensure that we always work with sparse matrices, that can be represented by row, column and value, only for non-zero entries. In this case, space complexity drops to $O(|E|)$, which for sparse graphs is usually a lot lower than the number of edges in a complete graph, i.e., $|E| \ll |V|^2$. This means that a graph with 10 million edges, assuming that the row, column and value each require 8 bytes to store, would only need $229$~MB of memory to be stored. Since matrix $H$ is sparse, we can simply allocate space for $H$ using a sparse matrix representation, for the zero-sum mask vector $a$, for the vector $e$ of ones, and for the PageRank vector $r$, and we can then compute matrix $\mathcal{M}$ on-the-fly during power iteration cycles as shown in Equation~\ref{eq:pr:sparse}.

\begin{equation}
  \label{eq:pr:sparse}
  r_{t+1} = \left( \alpha H + \left( \alpha a + (1 - \alpha) e \right) \left( \frac{1}{|V|} e^T \right) \right) r_t
\end{equation}

%***********************************************************************
\subsubsection{Fatigued PageRank computation using power iteration}
%***********************************************************************

Like PageRank, Fatigued PageRank also considers the idea of teleportation as a way to avoid sinks or cycles, leaving the corresponding term of the equation unchanged. Unlike PageRank, the navigation term is not only based on the outgoing transitions $H$, but also on a fatigue-derived factor. In particular, we approximate the probability of fatigue of a node based on its indegree vector $k^-$. We then normalize $k^-$ based on the maximum possible indegree $|V| - 1$ (ignoring loops), while using additive smoothing in order to avoid completely removing transitions to nodes with a maximum probability of fatigue. We use a low impact smoothing constant \mbox{$\beta = 0.1$}, mainly just to avoid zeros --- a zero would completely block a transition, while a small probability will provide a chance for the transition to have an effect during power iteration. Finally, we combine $H$ with the complement of the normalized and smoothed indegree, after renormalizing the vector to ensure the it remains stochastic. We do this through  element-wise multiplication ($\odot$) with each column $j$ of $H$, as show in Equation~\ref{eq:fpr:fatigue-decay}.

\begin{equation}
  \label{eq:fpr:fatigue-decay}
  \begin{split}
    k^* & = 1 - \frac{k^- + \beta}{|V| - 1 + \beta}\\
    H'_{\cdot j} & = \frac{k^*}{\left\lVert k^* \right\rVert_1} \odot H_{\cdot j}
  \end{split}
\end{equation}\\[-.5\baselineskip]

\noindent Essentially, a node becomes less probable to visit, if it has a high probability of getting fatigued. Fatigued PageRank is then computed based on power iteration, as defined in Equation~\ref{eq:pr}, after replacing $H$ with $H'$ in the computation of $S$ (Equations~\ref{eq:pr:S} and \ref{eq:pr:sparse}).

%***********************************************************************
\subsubsection{Exemplifying Fatigued PageRank computation}
%***********************************************************************

In this section, we illustrate the calculation of Fatigued PageRank, using the toy example graph in Figure~\ref{fig:toy-graph}. We prepared a graph with two sources (nodes 1 and 4 only have outgoing links) and a sink (node 5 only has incoming links). Sink nodes serve to illustrate the need for a teleportation term in PageRank (or Fatigued PageRank), while source nodes serve to illustrate the effect of minimum fatigue --- as we can see in $k^*$, both nodes 1 and 4 have maximum probability ($k_1^* = k_4^* = 0.26$). We also attempted to include a node with the maximum number of inlinks (i.e., $|V| - 1$ for a graph without loops). However, in order to keep the toy example clean and to ensure we had at least one sink, we opted to only include node 3 with a high indegree of $|V| - 2$. Notice, however, that a node with indegree $|V| - 1$ would result in a normalized indegree of one and therefore correspond to a zero entry in $k^*$ for $\beta = 0$ --- that is, without additive smoothing, transitions to nodes with maximum indegree would be completely blocked (we use $\beta = 0.1$).

\begin{minipage}{.45\linewidth}
  \begin{figure}[H]
    \centering
    \includegraphics[width=.7\linewidth,trim={80pt 80pt 60pt 60pt},clip]{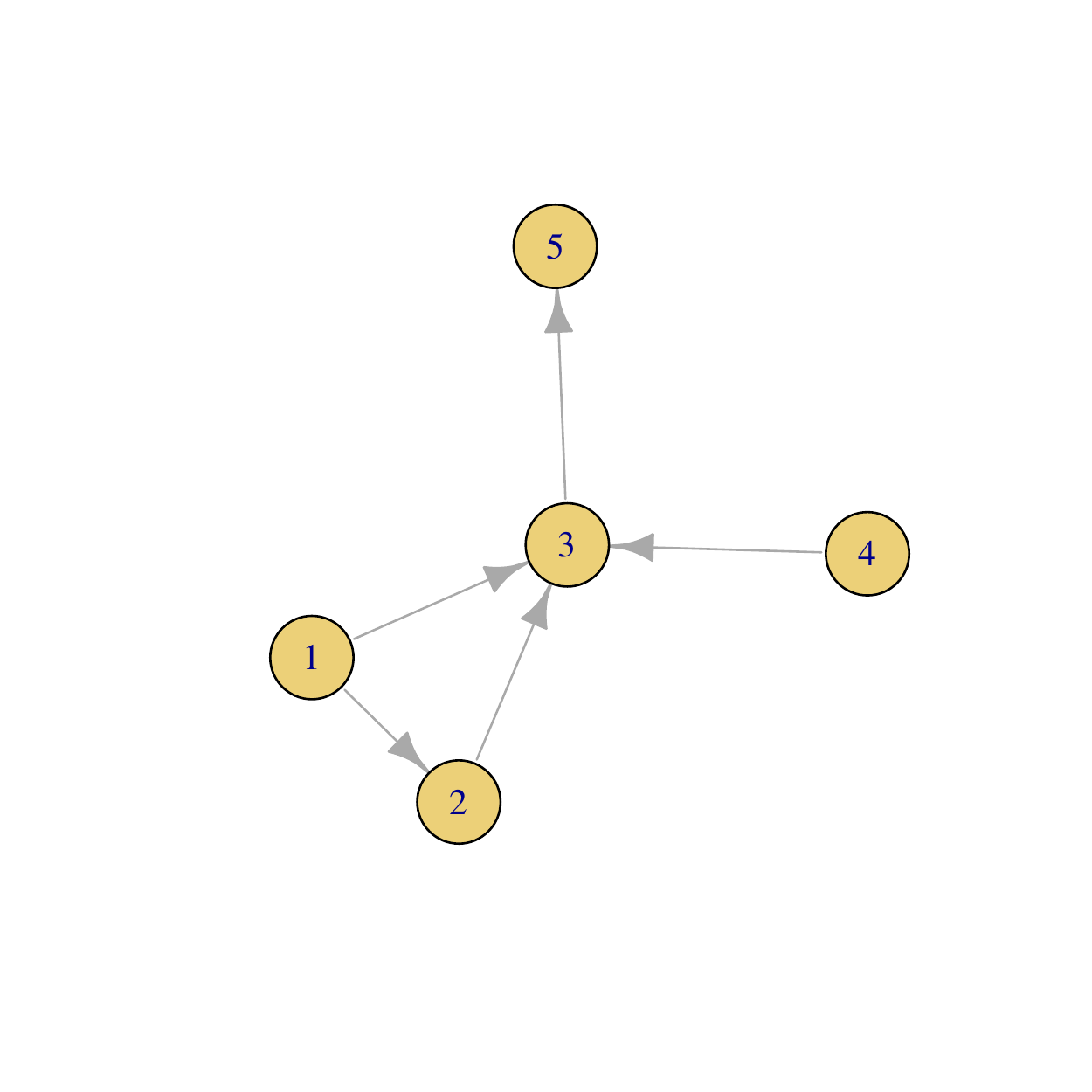}
    \caption{Toy graph, with one source (node 4) and one sink (node 5)}
    \label{fig:toy-graph}
  \end{figure}
\end{minipage}
\hfill
\begin{minipage}{.47\linewidth}
  \[
    \begin{aligned}
      A & = \begin{bmatrix}
        0 & 1 & 1 & 0 & 0 \\
        0 & 0 & 1 & 0 & 0 \\
        0 & 0 & 0 & 0 & 1 \\
        0 & 0 & 1 & 0 & 0 \\
        \bf 0 & \bf 0 & \bf 0 & \bf 0 & \bf 0 \\
      \end{bmatrix}\\
      H & = \begin{bmatrix}
        0.00 & 0.00 & 0.00 & 0.00 & \bf 0.00 \\
        0.50 & 0.00 & 0.00 & 0.00 & \bf 0.00 \\
        0.50 & 1.00 & 0.00 & 1.00 & \bf 0.00 \\
        0.00 & 0.00 & 0.00 & 0.00 & \bf 0.00 \\
        0.00 & 0.00 & 1.00 & 0.00 & \bf 0.00 \\
      \end{bmatrix}\\
      a^T & = \hspace{.3ex}\begin{bmatrix}{}
        \hspace{3ex} 0 & \hspace{2.8ex} 0 & \hspace{2.9ex} 0 & \hspace{2.9ex} 0 & \hspace{2.9ex} \bf 1 \\
      \end{bmatrix}\\
    \end{aligned}
  \]
\end{minipage}\\[1em]

As we can see, we begin with the adjacency matrix $A$, which we transform into $H$ by transposing and normalizing columns that are not all zeros. Zero-sum columns are then identified by a $1$ in the corresponding position of vector $a$. The corresponding row of $A$, column of $H$ and value of $a$ are all displayed in bold for a clearer understanding of such a process. While the zero-sum columns of $H$ could be have been replaced by the teleportation probability, we avoid doing so to save memory, taking better advantage of a sparse matrix representation. We instead do all computations on-the-fly during power iteration, resulting in a low memory footprint, since $H$ is static and only the PageRank vector must be updated (which can even be done in-place to further save memory). We can then set the damping factor to $\alpha = 0.85$, allocate a vector $e$ of ones, calculate the probability of teleportation $\frac{1}{|V|} = 0.20$ and calculate PageRank using Equation~\ref{eq:pr:sparse} by initializing $r_0$ for instance to the uniform probability $0.20$. For large graphs, the computation can even be done using blocks of rows of $H$, along with the corresponding elements of $a$ and $e$ ($e^T$ remains untouched, though). This results in incremental blocks of $r_t$ that can be sequentially concatenated and even processed in parallel.

\medskip

\begin{minipage}{.5\linewidth}
  \[
    \begin{aligned}
      \left[k^*\right]^T & = \hspace{.2ex} \begin{bmatrix}{}
        \hspace{.3ex} 0.26 & 0.20 & 0.08 & 0.26 & \hspace{.3ex} 0.20 \hspace{.3ex} \\
      \end{bmatrix}\\
      H' & = \begin{bmatrix}{}
        0.00 & 0.00 & 0.00 & 0.00 & \bf 0.00 \\
        0.71 & 0.00 & 0.00 & 0.00 & \bf 0.00 \\
        0.29 & 1.00 & 0.00 & 1.00 & \bf 0.00 \\
        0.00 & 0.00 & 0.00 & 0.00 & \bf 0.00 \\
        0.00 & 0.00 & 1.00 & 0.00 & \bf 0.00 \\
      \end{bmatrix}
    \end{aligned}
  \]
\end{minipage}
\hfill
\begin{minipage}{.45\linewidth}
  \[
    \begin{aligned}
      r_0^T & = \begin{bmatrix}{}
        0.20 & 0.20 & 0.20 & 0.20 & 0.20 \\
      \end{bmatrix}\\
      r_1^T & = \begin{bmatrix}{}
        0.03 & 0.15 & 0.42 & 0.03 & 0.37 \\
      \end{bmatrix}\\
      \ldots\\
      r_{10}^T & = \begin{bmatrix}{}
        0.05 & 0.09 & 0.23 & 0.05 & 0.59 \\
      \end{bmatrix}
    \end{aligned}
  \]
\end{minipage}

\medskip

For Fatigued PageRank, however, we also need to compute $k^*$, which will by multiplied by each column of $H$ to generate $H'$ as shown above. Power iteration will then incrementally update $r_t$ until convergence --- we show the values for $r_0$, $r_1$ and $r_{10}$ to illustrate how conversion happens in only $10$ iterations.

%***********************************************************************
\section{Evaluation}
\label{sec:eval}
%***********************************************************************

In this section, we assess the quality of Fatigued PageRank as a node ranking metric, based on the link graph for Simple English Wikipedia and using clickstream transitions as the ground truth. We then evaluate the effect of Fatigued PageRank as a query-independent feature in search, converting it to a relevance score that we use to rerank results from BM25. In both experiments, we always compare Fatigued PageRank with indegree, HITS authority and PageRank.

%***********************************************************************
\subsection{Fatigued PageRank as a node ranking metric}
\label{sec:eval:netsci}
%***********************************************************************

In order to assess the quality of Fatigued PageRank as a node ranking metric, we used an evaluation strategy similar to Dimitrov et al.~\cite{Dimitrov2017}, based on a Wikipedia's link graph, annotated with the number of transitions from its clickstream as the ground truth.

%***********************************************************************
\subsubsection{Simple English Wikipedia link graph}
%***********************************************************************

% CAMERAREADY replace with \cite{Devezas2019}
% [\footnote{Reference omitted for double blind review.}]
The Simple English Wikipedia Link Graph~\cite{Devezas2019} was built from the Simple English \texttt{page}\footnote{\scriptsize\url{https://dumps.wikimedia.org/simplewiki/20190101/simplewiki-20190101-page.sql.gz}} and \texttt{pagelinks}\footnote{\scriptsize\url{https://dumps.wikimedia.org/simplewiki/20190101/simplewiki-20190101-pagelinks.sql.gz}} SQL dumps from January 1st, 2019, using page names as node identifiers and considering only pages within the \texttt{article} namespace, with links from other pages. The SQL query used to prepare the graph was the following:
\begin{lstlisting}[language=sql,aboveskip=1.25em,belowskip=1.25em]
SELECT p1.page_title AS SOURCE, pl_title AS target
FROM pagelinks
JOIN page AS p1 ON pl_from = p1.page_id
WHERE pl_namespace = 0 AND pl_from_namespace = 0
\end{lstlisting}

\noindent The graph was then stored as a gzipped GML format. Given no clickstream data was directly available for the Simple English Wikipedia, we used the clickstream data for the English Wikipedia from December 2018\footnote{\scriptsize\url{https://dumps.wikimedia.org/other/clickstream/2018-12/clickstream-enwiki-2018-12.tsv.gz}}, adding a \texttt{transitions} attribute to each edge, that would be zero whenever information was unavailable. This resulted in a graph with $897,577$ nodes and $6,986,460$ edges, with an average outdegree of $7.78 \pm 49.52$ and an average indegree of $8 \pm 62$, based on information not only from existing pages at the date, but also from deleted pages that had been linked to existing pages. This justifies the direct usage of \texttt{pl\_title} instead of a \texttt{page\_id} attribute in table \texttt{pagelinks}, since these entries wouldn't have a corresponding entry in the \texttt{pages} table.

%***********************************************************************
\subsubsection{Measuring node rank correlation}
%***********************************************************************

\begin{table}[t]
  \centering
  \caption{Top 5 nodes according to each node ranking metric for Simple English Wikipedia.}
  \label{tab:top-simplewiki}

  \subfloat[Ranking by indegree and HITS authority.\label{tab:top-simplewiki:part1}]{
    \begin{tabularx}{.98\linewidth}{rXX}
      \toprule
      & \bf Indegree & \bf HITS Authority \\
      \midrule
        1 & \it United States & \it Lisa Bonet \\
        2 & \it France & \it Road to Paloma \\
        3 & \it International Standard Book Number & \it Ronon Dex \\
        4 & \it Geographic coordinate system & \it Native Hawaiians \\
        5 & \it Americans & \it Aquaman \\
      \bottomrule
    \end{tabularx}
  }

  \subfloat[Ranking by PageRank and Fatigued PageRank.\label{tab:top-simplewiki:part2}]{
    \begin{tabularx}{.98\linewidth}{rXX}
      \toprule
      & \bf PageRank & \bf Fatigued PageRank \\
      \midrule
        1 & \it United States & \it United States \\
        2 & \it United Kingdom & \it International Standard Book Number \\
        3 & \it India & \it France \\
        4 & \it List of United States cities by population & \it United Kingdom \\
        5 & \it Periodic table & \it City \\
       \bottomrule
    \end{tabularx}
  }
\end{table}

We computed the indegree, HITS authority, PageRank and Fatigued PageRank for the Simple English Wikipedia Link Graph. The top 5 pages according each metric are shown in Table~\ref{tab:top-simplewiki}. As we can see, all rankings are distinct, with ``United States'' being the only common entity across indegree, PageRank and Fatigued PageRank. On the other hand, HITS authority contains a notably different list of entities when compared to the remaining rankings, which is more related to entertainment, namely representing actors, movies and Hawaii (possible in reference to Jason Momoa, who plays the ``Ronon Dex'' character).

For each metric, we then computed its Pearson and Spearman correlations with the number of visits (i.e., the sum of incoming transitions). This follows the strategy used by Dimitrov et al.~\cite{Dimitrov2017} in their study of the Weighted PageRank, however we used a much smaller Wikipedia graph. We are particularly interested in the Spearman correlation, which compares rankings, but we also included the Pearson correlation as a way to characterize behavior. We analyzed the $897,577$ Simple English Wikipedia articles. The best overall results were obtained for PageRank, which achieved a Spearman correlation of $0.9902$, followed by HITS authority with $0.6330$, indegree with $0.3920$ and, only then, Fatigued PageRank with $0.1604$.

\begin{figure}[t]
  \centering
  \includegraphics[width=\linewidth]{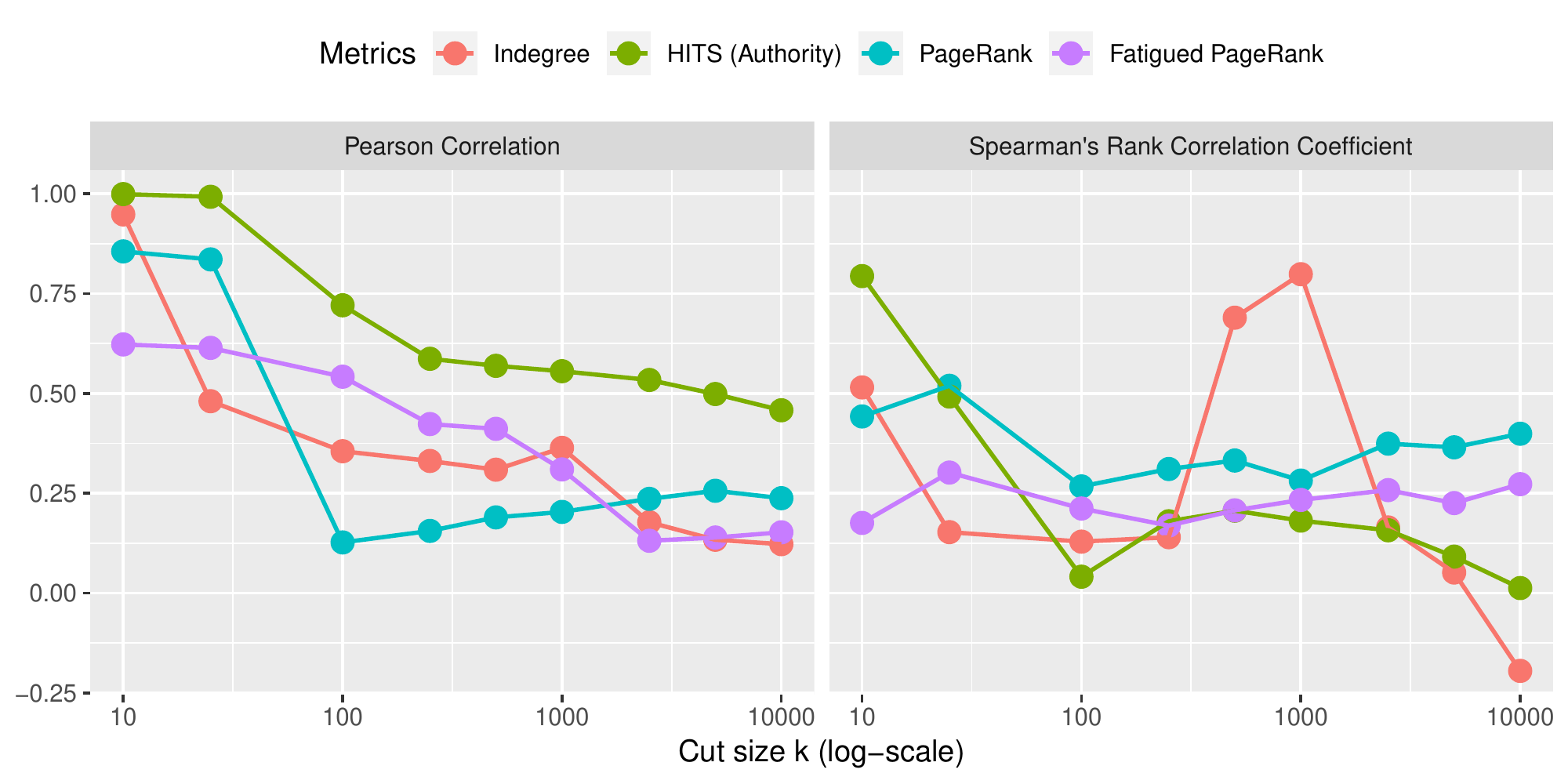}
  \caption{Comparison of node ranking metrics with the number of visits, given by the sum of incoming transitions from the Wikipedia clickstream dataset.}
  \label{fig:netsci-eval}
\end{figure}

The low results for Fatigued PageRank led us to investigate further, by looking at the correlation for different ranking cuts of size $k$, according to each metric, as illustrated in Figure~\ref{fig:netsci-eval}. In particular, we plotted the correlation coefficients for $k \in \{ 10, 25, 100, 250, 500, 1000, 2500, 5000, 10000 \}$ using a log-scale. This means that, for instance for PageRank, we sorted from highest to lowest PageRank and kept the top $k$ values, which we then correlated with the corresponding number of visits. When looking at Spearman correlation, we found that, for the top 10, we obtained the best results using HITS authority. We also found that, for the top 10, the indegree is a good approximation of PageRank, which is consistent with the literature~\cite{Fortunato2006}. As the cut size increases, though, the quality of HITS authority decreases substantially (it had the lowest coefficient for the top 100), remaining low but fairly stable until the top 10000. Another interesting characteristic we found was the high variability of the quality of the indegree as a node ranking metric.

As we can see in the figure, the Spearman correlation for the indegree suddenly increases at around top 1000, only to sharply drop after that. For that cut, we found that there were several ties where the indegree for most nodes was around $705$, while the corresponding number of visits was also tied with values around zero --- this was responsible for the sudden boost in correlation. However, when further ahead the wrongly ranked nodes are added to the top, the correlation sharply drops to a negative coefficient. This variability shows the ``naïveness'' of the indegree.

\begin{table}
  \centering
  \caption{Variance of the correlations over all cuts, in ascending order by the variance of Spearman's rank correlation.}
  \label{tab:cor-var-simplewiki}

  \begin{tabularx}{\linewidth}{XCC}
    \toprule
    \multirow{2}{*}{\bf Metric} & \multicolumn{2}{c}{\bf Variance}\\
    & \bf Pearson & \bf Spearman \\
    \midrule
    Fatigued PageRank & 0.0399 & 0.0019 \\
    PageRank          & 0.0826 & 0.0064 \\
    HITS (Authority)  & 0.0421 & 0.0624 \\
    Indegree          & 0.0631 & 0.1052 \\
    \bottomrule
  \end{tabularx}
\end{table}

Out of all the metrics, only PageRank and Fatigued PageRank have shown to be consistently stable for different cut sizes, however Fatigued PageRank appears to be consistently lower than PageRank. This can be explained by the fact that, inherently, fatigue will take away from visiting the most popular nodes. We hypothesize that fatigue introduces diversity and somewhat ``redistributes the wealth'', potentially providing a better way to explore the graph, by reducing the bias of popularity. This, however, requires further investigation and is out of the scope of this paper.

As shown in Table~\ref{tab:cor-var-simplewiki}, we inspected the variance of the overall correlation values, in order to understand the consistency of each metric for increasing cut sizes. We found that Fatigued PageRank obtained the lowest variances for the Pearson and Spearman's rank correlations, showing robustness to varying the cut size. When considering Pearson's correlation, PageRank obtained the highest (worst) variance, while, for Spearman's rank correlation, it obtained the second lowest (best) variance. On the other side, as expected, the indegree had the highest (worst) variance in both cases, illustrating its lack of robustness.

%***********************************************************************
\subsection{Fatigued PageRank as a query-independent feature in search}
\label{sec:eval:ir}
%***********************************************************************

 In this section, we evaluate the impact of Fatigued PageRank in retrieval effectiveness, when combined with BM25, in comparison with other graph-based metrics. These metrics were computed over a companion link graph built from TREC Washington Post Corpus, using the inter-document hyperlinks found in HTML anchors.

 %***********************************************************************
\subsubsection{TREC Washington Post Corpus}
%***********************************************************************

The TREC Washington Post Corpus\footnote{\url{https://trec.nist.gov/data/wapost/}} was first used during TREC 2018 in the Common Core and News tracks. It contains $608,180$ documents collected from the Washington Post between January 2012 and August 2017. Out of these documents, we only used the subset of $595,037$ documents ($236,649$ news articles and $358,388$ blog posts), that resulted of the removal of duplicates by \texttt{id}. Multiple duplicates by \texttt{title} still exist in the collection, however the TREC Common Core track evaluation data that is available was prepared over this subset. In order to build the link graph, we translated each hyperlink into the corresponding document \texttt{id}, which acted as a node identifier, otherwise ignoring the hyperlink when it didn't point to a valid document in the corpus. This resulted in a graph with $159,228$ nodes and $319,985$ edges, with an average outdegree of $2.01 \pm 2.63$ and an average indegree of $2.01 \pm 7.87$. For news articles, the average outdegree was $1.43 \pm 2.28$ and the average indegree was $2.39 \pm 9.57$, while, for blog posts, the average outdegree was $2.22 \pm 2.71$ and the average indegree was $1.87 \pm 7.15$. Overall, there was a lower number of outgoing links departing from news articles. On the other hand, news articles also received the highest number of incoming links.

%***********************************************************************
\subsubsection{Measuring retrieval performance impact}
%***********************************************************************

\begin{table}[t]
  \centering
  \caption{Top 5 nodes according to each node ranking metric for TREC Washington Post Corpus.}
  \label{tab:top-wapo}

  \subfloat[Ranking by indegree and HITS authority.\label{tab:top-wapo:part1}]{
    \begin{tabularx}{\linewidth}{rXX}
      \toprule
      & \bf Indegree & \bf HITS Authority \\
      \midrule
      1 & \it Trump recorded having extremely lewd conversation about women in 2005 & \it This photo of an officer comforting a baby went viral. But there’s more to the story. \\
      2 & \it This professor has predicted every presidential election since 1984. He’s still trying to figure out 2016. & \it ‘Heartbreaking’ video captures toddler trying to wake mother after apparent overdose \\
      3 & \it This photo of an officer comforting a baby went viral. But there’s more to the story. & \it Ohio city shares shocking photos of adults who overdosed with a small child in the car \\
      4 & \it Trump is headed for a win, says professor who has predicted 30 years of presidential outcomes correctly & \it The heroin epidemic’s toll: One county, 70 minutes, eight overdoses \\
      5 & \it ‘Heartbreaking’ video captures toddler trying to wake mother after apparent overdose & \it At 18 years old, he donated a kidney. Now, he regrets it. \\
      \bottomrule
    \end{tabularx}}

  \subfloat[Ranking by PageRank and Fatigued PageRank.\label{tab:top-wapo:part2}]{
    \begin{tabularx}{\linewidth}{rXX}
      \toprule
      & \bf PageRank & \bf Fatigued PageRank \\
      \midrule
      1 & \it Trump recorded having extremely lewd conversation about women in 2005 & \it Here’s a guide to the sex allegations that Donald Trump may raise in the presidential debate \\
      2 & \it 78 Republican politicians, donors and officials who are supporting Hillary Clinton & \it The facts about Hillary Clinton and the Kathy Shelton rape case \\
      3 & \it An unlikely Bush finally did some damage to Donald Trump: Billy Bush & \it Khizr Khan’s loss: A grieving father of a soldier struggles to understand \\
      4 & \it Professor who predicted 30 years of presidential elections correctly called a Trump win in September & \it The father of Muslim soldier killed in action just delivered a brutal repudiation of Donald Trump \\
      5 & \it Here’s a guide to the sex allegations that Donald Trump may raise in the presidential debate & \it Obamas sign book deals with Penguin Random House \\
      \bottomrule
    \end{tabularx}}
\end{table}

We computed the indegree, HITS authority, PageRank and Fatigued PageRank for the TREC Washington Post Corpus link graph. The top 5 documents (news articles or blog posts) according to each metric are shown in Table~\ref{tab:top-wapo}. The behavior is similar to the rankings for the Simple English Wikipedia (cf.~Section~\ref{sec:eval:netsci}), where each metric results in a distinct ranking. The indegree and HITS authority share two common titles, while the indegree and PageRank, as well as PageRank and Fatigued PageRank, share one common title. Once again, HITS authority contains a large fraction of potentially-viral titles, showing its ability to rank from an entertainment point of view --- in Wikipedia we had found movies and actors in the top ranks according to HITS authority.

\begin{table}[!t]
  \centering
  \caption{Retrieval effectiveness of graph-based metrics, as query-independent evidence, when combined with BM25 using the \emph{sigm} function.}
  \label{tab:effectiveness}

  \begin{tabularx}{\linewidth}{Xllcr}
    \toprule
    \bf Model & \bf GMAP & \bf MAP & \bf NDCG@10 & \bf P@10\\
    \midrule
    BM25                      & 0.1395 & 0.2031 & 0.3528 & 0.3700\\
    \midrule
    BM25 + Indegree           & 0.1357 & 0.1994 & 0.3537 & 0.3800\\
    BM25 + HITS Authority     & 0.1395 & 0.2031 & 0.3540 & 0.3720\\
    BM25 + PageRank           & 0.1395 & 0.2031 & 0.3528 & 0.3700\\
    BM25 + Fatigued PageRank  & 0.1395 & 0.2031 & 0.3528 & 0.3700\\
    \bottomrule
  \end{tabularx}
\end{table}

In search, we can use graph-based features as query-independent evidence (e.g., PageRank) that can be combined with other retrieval models (e.g., BM25). This can be done for instance through a linear combination, as a feature in a learning-to-rank model, or through a post-processing reranking approach. We followed the strategy by Craswell et al.~\cite{Craswell2005}, described in Section~\ref{sec:ref-work:search}, for transforming graph-based evidence into a relevance score, which can then be combined with a text-based relevance score by adding both terms. This enabled us to rerank BM25 results using a graph-based feature. In particular, we applied the \emph{sigm} function to each graph-based metric, with parameters $w = 1.8$, $k = 1$ and $a = 0.6$, which had achieved the best results in the work by Craswell et al. This enabled us to evaluate Fatigued PageRank's impact in retrieval effectiveness, when compared to alternative metrics.

Table~\ref{tab:effectiveness} shows the results for this evaluation, where we used BM25 as the baseline and computed classic metrics, like MAP, NDCG@10 and P@10, for measuring effectiveness. We also included GMAP, which is less sensitive to outliers than MAP, since it uses the geometric mean instead of the arithmetic mean to aggregate the average precisions --- when GMAP is lower than MAP, it usually means that only a few topics were driving up the score, despite most of them actually having a lower average precision than MAP would lead us to believe. The results show that the overall impact of the graph-based metrics is minimal, except for the indegree, which decreases effectiveness for GMAP and  MAP, but increases effectiveness for NDCG@10 and P@10, which only consider the top 10 results. This is consistent with the results shown in Figure~\ref{fig:netsci-eval}, where the indegree is shown to be slightly better than PageRank, but only for the top 10.

\begin{table}[!t]
  \centering
  \caption{Top 5 most frequent document titles in TREC Washington Post Corpus.}
  \label{tab:wapo-duplicates}

  \begin{tabularx}{\linewidth}{rX}
    \toprule
    \bf Freq. & \bf Title\\
    \midrule
    455 & \it Happy Hour Roundup \\
    353 & \it A 7-year-old told her bus driver she couldn’t wake her parents. Police found them dead at home. \\
    320 & \it How long before the white working class realizes Trump was just scamming them? \\
    310 & \it Five dead teens, a stolen cop car and the ‘most horrific’ crash in decades \\
    252 & \it Trump is headed for a win, says professor who has predicted 30 years of presidential outcomes correctly \\
    \bottomrule
  \end{tabularx}
\end{table}

The results were quite neutral regarding the overall benefits of graph-based metrics to improve retrieval effectiveness over the TREC Washington Post Corpus. We hypothesize that, despite their unique identifiers and URLs, the high number of pages with a duplicate title (cf.~Table~\ref{tab:wapo-duplicates}) affected both the text-based relevance score and the graph-based relevance score. It introduced noise that, by being removed, could have provided a better insight into the effect of the studied query-independent evidence. This would, however, require a different set of relevance judgments that are not currently available.

%***********************************************************************
\section{Conclusions}
\label{sec:concl}
%***********************************************************************

Inspired by the response of neurons in the brain when stimulated, we proposed the idea of fatigue as a strategy for random walking in graphs. We applied this idea to PageRank, formalizing a novel Fatigued PageRank metric that follows a random explorer model, being analogous to the combination of PageRank and Reverse PageRank, or the authority and hub scores from HITS. We then evaluated our graph-based metric, comparing it with the indegree, HITS authority and PageRank, when computed over the Simple English Wikipedia link graph. We analyzed the rankings and, according to Spearman's rank correlation coefficient, assessed the quality of each metric in comparison with the number of visits (provided by the sum of incoming transitions, according to the English Wikipedia clickstream data). We found that, while Fatigued PageRank obtained the lowest overall correlation of $0.1604$, it was actually able to outperform the indegree and HITS authority for the top $10,000$ nodes, showing a more consistent behavior than both those metrics, with the lowest variance for the Spearman's rank correlations of all metrics. Despite having a lower overall correlation score than PageRank, the fatigued version might work in favor of bias reduction towards the most popular nodes, ensuring diversity in the preparation of the ranking.

\sloppy We also assessed the impact in retrieval effectiveness of Fatigued PageRank when used as query-independent evidence, in combination with a text-based relevance score like BM25. We used the sigmoid function proposed by Craswell et al.\ to transform the graph-based feature into a static relevance weight that could be used as an additional signal during ranking. Based on the TREC Washington Post Corpus, we were unable to find significant differences in the retrieval effectiveness, except for the indegree metric, which decreased GMAP and MAP, but increased NDCG@10 and P@10, over a BM25 baseline. We believe the existence of duplicate documents in the collection was detrimental to this experiment, given the positive results with graph-based metrics found in the literature (e.g., Najork~\cite[Figure 3]{Najork2007}). Another possible explanation would be regarding the specificity of the queries used for evaluation, given that general queries usually benefit from graph-based metrics, while specific queries do not (cf.\ Najork~\cite[Figure 4]{Najork2007}).

\subsection{Future work}

In the future, we would like to reiterate over the computation approach of the Fatigued PageRank, namely exploring a more approximate analogy to the combination of PageRank and Reverse PageRank, replacing the $k^*$ vector with the Markov matrix used in Reverse PageRank and studying the differences. We would also like to further characterize the specificity of TREC Washington Post Corpus queries, in order to better understand what to expect from the impact that graph-based metrics should have in effectiveness. As an alternative, we would like to repeat the experiment with a different dataset, in order to validate the results based on the existence or lack of consistency. Finally, we would like to explore alternative ways to combine the graph-based features with a text-based relevance weight, namely exploring the remaining functions proposed by Craswell et al., along with different parameter configurations, as well as other approaches, like the linear combinations proposed by Najork.

\begin{acks}
  José Devezas was supported by research grant PD/BD/128160/2016, provided by the Portuguese national funding agency for science, research and technology, Fundação para a Ciência e a Tecnologia (FCT), within the scope of Operational Program Human Capital (POCH), supported by the European Social Fund and by national funds from MCTES.
\end{acks}

\bibliographystyle{ACM-Reference-Format}
\bibliography{fatigued-pagerank}

\end{document}